# Graphene field-effect transistors based on boron nitride gate dielectrics

Inanc Meric[1], Cory R. Dean[1,3], Andrea F. Young[2], James Hone[3], Philip Kim[2], and Kenneth L. Shepard[1]

[1]Department of Electrical Engineering, [2]Department of Physics, [3]Department of Mechanical Engineering, Columbia University, New York, NY, 10027
Tel: (212) 854-2529, Fax: (212) 932-9421, Email: shepard@ee.columbia.edu

## Abstract

Graphene field-effect transistors are fabricated utilizing single-crystal hexagonal boron nitride (h-BN), an insulating isomorph of graphene, as the gate dielectric. The devices exhibit mobility values exceeding 10,000 cm$^2$/V-sec and current saturation down to 500 nm channel lengths with intrinsic transconductance values above 400 mS/mm. The work demonstrates the favorable properties of using h-BN as a gate dielectric for graphene FETs.

## Introduction

One of the major obstacles to the development of graphene field-effect transistors (GFETs) remains engineering the dielectric interfaces to the graphene. With the exception of "suspended" graphene samples[1], which are not practical for devices, most GFETs are fabricated on silicon dioxide substrates and top-gated with high-κ gate dielectrics grown on the graphene. The deposition of these top gate dielectrics often involves the initial deposition of a noncovalent functionalization layer absorbed on the graphene surface.[2] Both the top-gate oxide and supporting oxide substrate significantly degrade the electronic properties of the graphene. Charged impurities trapped in the dielectrics or at the graphene-dielectric interfaces dope the graphene, may significantly degrade mobility, and can result in hysteretic current-voltage characteristics. Surface polar optical phonons from the substrate limit room temperature transport [3] and achievable saturation velocities [4]. In this paper, we present the first measurements of GFETs utilizing hexagonal boron nitride (h-BN) as both a gate dielectric and supporting substrate, resulting in dramatically improved transistor current voltage characteristics.

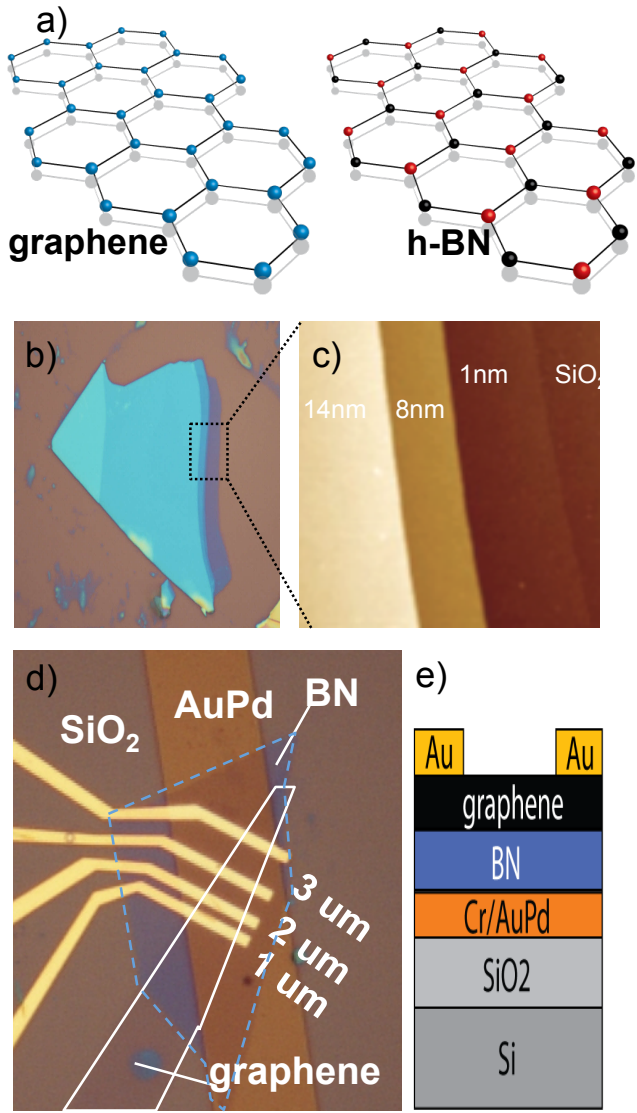

**Figure 1. Back-gated GFET with h-BN gate dielectric**. (a) Atomic structure of graphene and hexagonal boron nitride (b) optical micrograph of exfoliated h-BN (c) AFM image of h-BN showing different layer thickness (d) optical image of GFET; (e) schematic of the back-gated device structure in (d)

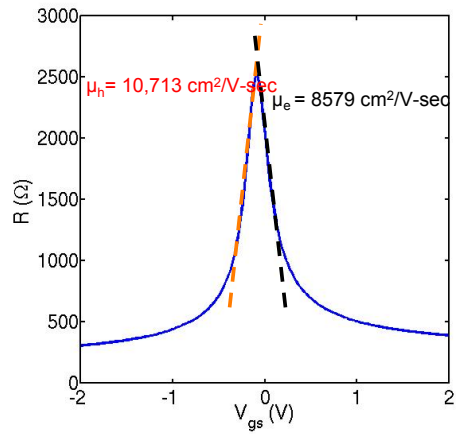

**Figure 2. Low-field transport characteristics of GFET device**. R=1/$g_{ds}$ at $V_{ds}$ = 10 mV as a function of $V_{gs}$ for W/L=3.4mm/2.8mm.

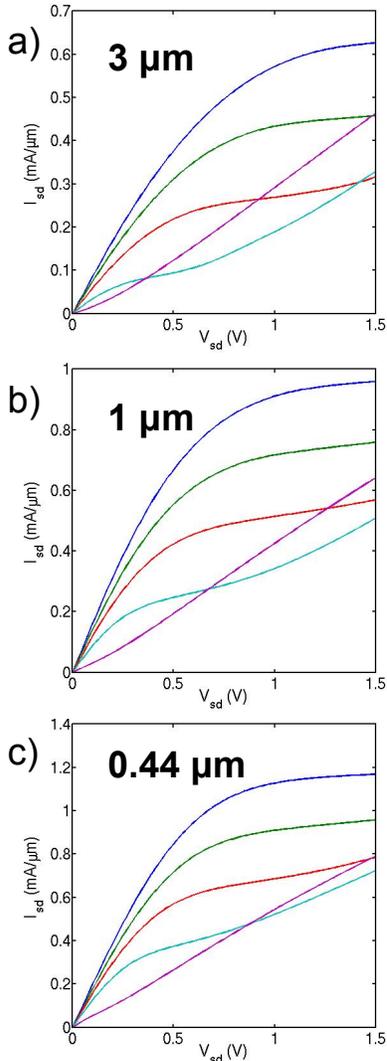

**Figure 3. Current-voltage characteristics of GFET devices.** (a) 3 μm channel length, (b) 1 μm channel length, (c) 0.44 μm channel length for $V_{gs}$ -2V to 0V in 0.5V steps

**Basic device structure**

h-BN is an insulating isomorph of graphite (Fig. 1) with boron and nitrogen atoms occupying the inquivalent A and B sublattices in the Bernal structure. The hexagonal structure with strong in-plane bonding makes the surface chemically inert and free of dangling bonds and surface charge traps. The surface is also atomically flat over large areas. The bandgap (5.97 eV) and dielectric properties of h-BN ($\varepsilon$~3-4 and $V_{Breakdown}$ ~ 0.7 V/nm) compare favorably with $SiO_2$. The excellent thermal conductivity of h-BN, 600 times higher then silicon dioxide, is also advantageous for FET applications to minimize device heating.

To fabricate graphene-on-BN, we employ a mechanical transfer process in which h-BN layers are exfoliated from ultra-pure h-BN single crystals and transferred onto predefined metal gates (1nm Cr/ 20nmAuPd).[5] The devices shown in Fig. 1 and measured here have a dielectric thickness of approximately 8.5 nm. Because the h-BN can be made arbitrarily thin (down to a single monolayer) our device geometry allows us to utilize the same h-BN dielectric layer as both a supporting substrate and local-gate dielectric. This, therefore, allows us to fabricate the required FET structure without an additional top gate. Cr/Au (1nm/90nm) electrodes are used as Ohmic contacts, producing p-type doping of the graphene under the contacts because of work-function differences. After transfer, the graphene on h-BN (as measured by atomic force microscopy) is approximately three times less rough than on $SiO_2$.

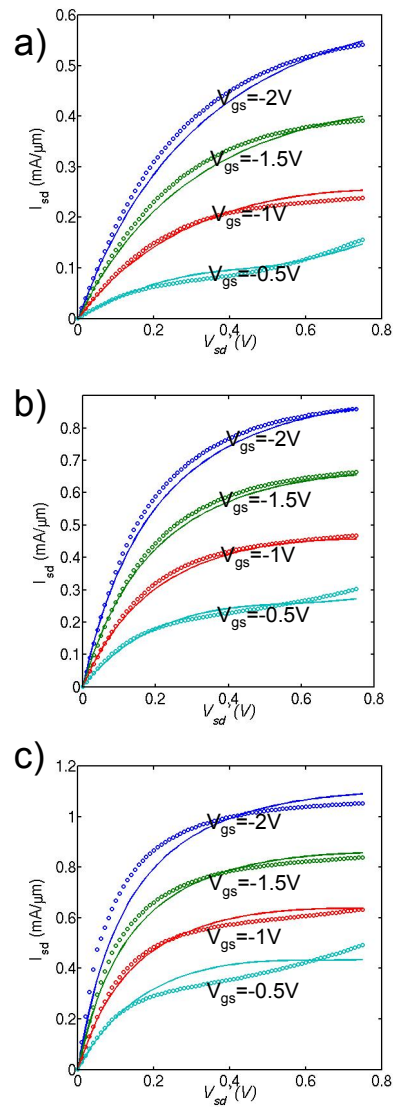

**Figure 4. Intrinsic device IV characteristics.** Intrinsic IV curves after the contact resistance extraction from the measured curves of Fig. 3. Channel length of (a) 3 μm; (b) 1 μm; and (c) 0.44 μm. Model fits (solid lines) are shown along with measured data.

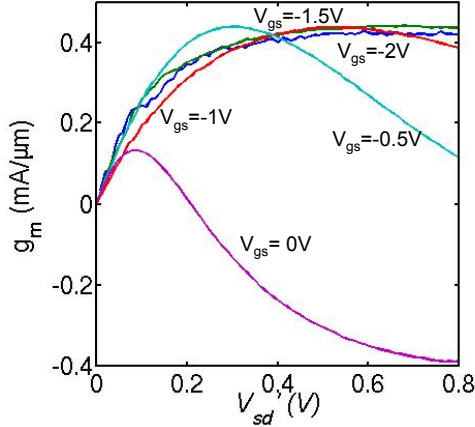

**Figure 5. Intrinsic small-signal transconductance ($g_m$)** as a function of drain-to-source voltage ($V_{sd}$) for 0.44 μm channel length

### Low-field transport

Fig. 2 shows the channel resistance ($1/g_{ds}$) at $V_{ds} = 10\ mV$ as a function of $V_{gs}$ for a $W/L = 3.4\ \mu m/2.8\ \mu m$ device. Low-field mobility of these devices exceeds *10,000 $cm^2$/V-sec*, as extracted from fits to the low-field transport measurements. The carrier concentration in the channel is given by Eq. 1 (Fig. 7(b)) where $n_0$ is the minimum sheet carrier concentration as determined by disorder and thermal excitation. Here, $n_0$ is approximately $2.2 \times 10^{11}\ cm^{-2}$. $C_g$ ($\cong 363\ nF/cm^2$) is given by the parallel combination of the electrostatic capacitance of the gate and the quantum capacitance of graphene (which ultimately limits achievable gate capacitances). There is almost no doping of the graphene channel with the location of the Dirac point given by $V_0 \cong -0.07\ V$ and with a gate-voltage hysteresis of less then *10 mV* at room temperature. BN-supported devices appear to be more stable compared to their $SiO_2$-supported counterparts, as heating and high-bias stress have virtually no effect on the transport characteristics. For a p-type channel (matching the doping of the source and drain contacts), the contact resistance is approximately *673 $\Omega/\mu m^2$*. n-type channels show a contact resistance that is approxi-mately 31 % higher for this device geometry because of the additional resistance of the p-n junction at the source/drain; therefore, subsequent large-signal characterization are performed with these devices as pFETs.

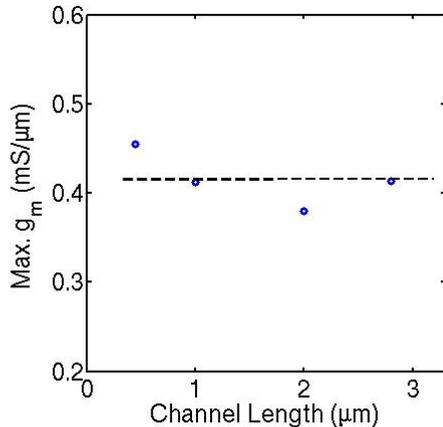

**Figure 6. Maximum intrinsic transconductance ($g_m$)** as a function channel length. The dashed line shows the average value of 415 mS/mm

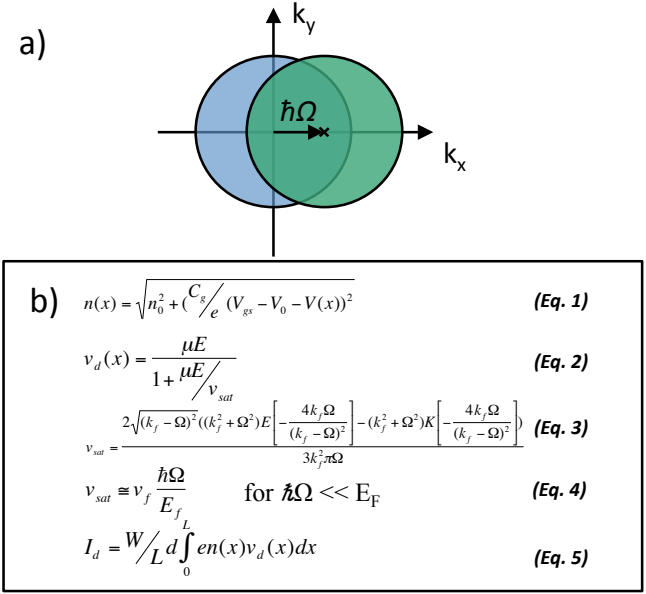

**Figure 7. GFET Modeling**. (a) shows the the nonequilibrium Fermi surface used for the modeling (b) equations used for field-effect modeling. E(k) is the complete elliptic integral of the second kind; K(k) is the complete elliptic integral of the first kind

### Current-voltage characteristics

Fig. 3 shows the I-V characteristics measured from devices with channel lengths of 3 μm, 1 μm, and 0.44 μm. In the unipolar regime, $V_{sd} < V_{sd\text{-}kink}$, the GFETs show saturating I-V characteristics, where $V_{sd\text{-}kink}$ is the drain bias at which the Dirac point enters the channel.[4] Because these devices are still limited by contact resistance, intrinsic device IV characteristics are shown in Fig. 4 after the extraction of the measured contact resistance. The 0.44-μm-channel-length device shows an intrinsic $I_{on}$ of more than *1 mA/μm*. Fig. 5 shows the associated intrinsic transconductance for this same 0.44-μm-channel-length device as a function of $V_{sd}$ for different values of $V_{gs}$. The peak intrinsic transconductance, obtained after the extraction of the contact resistance exceeds *400 mS/mm* and is independent of channel length as plotted in Fig. 6, consistent with velocity-saturation-dominated transport. This value is approximately 2.6 times higher than previously reported values on $SiO_2$-supported samples [4], even though the effective gate capacitance is 30 % lower.

### Device modeling and velocity saturation

Fig. 7 outlines the basic field-effect modeling of the devices (these model fits are shown in Fig. 4). The carrier-dependent saturation velocity ($v_{sat}$) which assumes a simple nonequilibrium Fermi surface shown in Fig. 7a is given by the expression in Eq. 3, and approaches $v_F$ for $\hbar\Omega\gg E_F$. Furthermore, we explicitly include the density-dependence of $v_{sat}$ self-consistently in the current-voltage model, which was not done in previous analyses.[4] Fig. 8 shows $v_{sat}$ as a function of $1/n^{1/2}$ (for an overdrive sufficiently large to ensure a unipolar channel) where $n$ is taken at the drain of the channel for three different channel lengths. $v_{sat}$ exceeds *1.14 x 10$^7$ cm/sec* at sheet densities of more than *4.5 x 10$^{12}$ cm$^{-2}$*, more than two times higher than results on SiO$_2$-supported devices with high-$\kappa$ gate dielectrics. [4] The slope of the curves indicates the optical phonon energy of approximately *40 meV*, significantly less than the surface polar optical phonon energy of *100 meV* for BN. Such lower energies have been observed consistently for GFET devices at high densities[6] and remain the subject of active investigation.

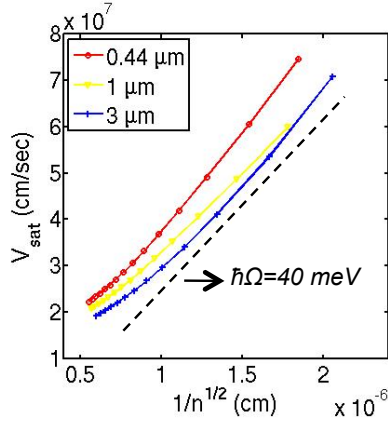

**Figure 8. Saturation velocity** plotted versus square root of the drain density for different channel lengths as extracted from the model fits. The dashed line shows a slope corresponding to $\hbar\Omega=40$ meV

**Acknowledgments**

The authors would like to thank K. Watanabe and T. Taniguchi for supplying h-BN crystals. The authors acknowledge the support of the C2S2 Focus Center, one of six research centers funded under the Focus Center Research Program (FCRP), a Semiconductor Research Corporation entity, and DARPA under contract FA8650-08-C-7838 through the CERA program, and by the AFOSR MURI Program on new graphene materials technology, FA9550-09-1-0705.

**References**

[1] K. I. Bolotin, K. J. Sikes, Z. Jiang, M. Klima, G. Fudenberg, J. Hone, P. Kim, and H. L. Stormer, "Ultrahigh electron mobility in suspended graphene," *Solid State Communications,* vol. 146, pp. 351-355, 2008.

[2] D. B. Farmer, H.-Y. Chiu, Y.-M. Lin, K. A. Jenkins, F. Xia, and P. Avouris, "Utilization of a buffered dielectric to achieve high field-effect carrier mobility in graphene transistors," in *Nano Lett.* vol. 9, 2009, pp. 4474-8.

[3] J.-H. Chen, C. Jang, S. Xiao, M. Ishigami, and M. S. Fuhrer, "Intrinsic and extrinsic performance limits of graphene devices on SiO2," *Nat Nano,* vol. 3, pp. 206-209, 2008.

[4] I. Meric, M. Y. Han, A. F. Young, B. Ozyilmaz, P. Kim, and K. L. Shepard, "Current saturation in zero-bandgap, top-gated graphene field-effect transistors," *Nature Nanotechnology,* vol. 3, pp. 654-659, 2008.

[5] C. R. Dean, A. F. Young, I. Meric, C. Lee, L. Wang, S. Sorgenfrei, K. Watanabe, T. Taniguchi, P. Kim, K. L. Shepard, and J. Hone, "Boron nitride substrates for high-quality graphene electronics," in *Nature Nanotechnology*, 2010.

[6] M. Freitag, M. Steiner, Y. Martin, V. Perebeinos, Z. Chen, J. Tsang, and P. Avouris, "Energy Dissipation in Graphene Field-Effect Transistors," in *Nano Lett*, 2009.